\documentclass[twocolumn,english,aps,pra,showpacs]{revtex4-1}
\usepackage[T1]{fontenc}
\usepackage[latin9]{inputenc}
\usepackage{color}
\usepackage{babel}
\usepackage{float}
\usepackage{amsmath}
\usepackage{graphicx}
\usepackage[unicode=true,pdfusetitle,
 bookmarks=true,bookmarksnumbered=false,bookmarksopen=false,
 breaklinks=false,pdfborder={0 0 0},backref=false,colorlinks=true]
 {hyperref}

\makeatletter
\@ifundefined{textcolor}{}
{%
 \definecolor{BLACK}{gray}{0}
 \definecolor{WHITE}{gray}{1}
 \definecolor{RED}{rgb}{1,0,0}
 \definecolor{GREEN}{rgb}{0,1,0}
 \definecolor{BLUE}{rgb}{0,0,1}
 \definecolor{CYAN}{cmyk}{1,0,0,0}
 \definecolor{MAGENTA}{cmyk}{0,1,0,0}
 \definecolor{YELLOW}{cmyk}{0,0,1,0}
}

\makeatother

\begin{document}

\title{Nondestructive verification of continuous-variable entanglement}

\author{Alencar J. de Faria}

\email{alencar.faria@unifal-mg.edu.br}

\affiliation{Instituto de Ciência e Tecnologia, Universidade Federal de Alfenas,
CEP 37715-400, Poços de Caldas, MG, Brazil}
\begin{abstract}
An optical procedure in the context of continuous variables to verify
bipartite entanglement without destroying both systems and their entanglement
is proposed. To perform the nondestructive verification of entanglement,
the method relies on beam-splitter and quantum nondemolition (QND)
interactions of the signal modes with two ancillary probe modes. The
probe modes are measured by homodyne detections, and the obtained
information is used to feedforward modulation of signal modes, concluding
the procedure. Characterizing the method by figures of merit used
in QND processes, we can establish the conditions for an effectively
quantum scheme. Based on such conditions, it is shown that the classical
information acquired from the homodyne detections of probe modes is
sufficient to verify the entanglement of the output signal modes.
The processing impact due to added noise on the output entanglement
is assessed in the case of Gaussian modes. 
\end{abstract}


\maketitle

\section{Introduction}

Entanglement is one of the most fundamental resources for performing
processes in quantum information and computation. Besides the technological
possibilities, the entanglement challenges our understanding of the
quantum world and its connection with classical physics \cite{Nielsen,Horodecki09,Kimble}.
Recently, many experiments have accomplished quantum communication
protocols sending light signals over distances of hundreds of kilometers
\cite{Yin12,Ma12,Inagaki13}. In these experiments, the entanglement
was an essential part. It has also been studied as the use of entangled
signals over long distances may increase the applicability of quantum
cryptography protocols \cite{Scheidl09}. Thus it is very natural
to devise stages along the transmission of entangled signals that
verify if such signals are really entangled, without destroying or
excessively disturbing them during the verification processes. In
other words, for future quantum communications, nondestructive certification
protocols of entanglement will be required, ensuring the use of entanglement
for subsequent processes of quantum information. Studies of nondestructive
entanglement verification or analysis have been done in the framework
of discrete variable systems, such as single photons \cite{Barrett05,Tang12,Liu16}. 

Differently from the previously cited studies, quantum communications
using entangled signals may also be carried out in the context of
continuous-variable systems, e. g., bright beams. The light beams
may be regarded as oscillation modes, such that the states are vectors
of an infinite-dimensional Hilbert space and observables are continuous
spectrum operators, analogously to the position and momentum operators
of the quantum harmonic oscillator \cite{Braunstein05}. For light
beams, we consider amplitude and phase operators, also called quadratures.
The research field of continuous-variable systems is very active and
has extensive literature. Examples of quantum information protocols
performed with continuous-variable light modes are quantum teleportation
\cite{Braunstein98,Furusawa98,Bowen03,Zhang03}, cloning \cite{Buzek96,Grover/Cirac99,Andersen05},
and telecloning \cite{Murao99,Loock01,Koike06}. In such cases, the
entanglement is an essential ingredient, therefore its conservation
and its verification are primordial to further more complex applications. 

Thus a minimally invasive measurement method to observe the entanglement
is desirable. Although all quantum measurement entails a back-action
effect, we can measure the signal, in order to preserve some of the
properties of its original state. In particular, a quantum nondemolition
(QND) measurement is able to measure an observable without disturbing
it, at the expense of a back-action disturbance on the conjugate observable
\cite{Braginsky80,Braginsky96,Grangier98,Wiseman-Milburn}. In quantum
optics, the QND measurements were initially performed by coupling
the signal and probe modes in nonlinear optical media, such as Kerr
media, optical fibers (third-order nonlinearity) \cite{Milburn83,Imoto85,Levenson86},
and in optical parametric amplifiers (second-order nonlinearity) \cite{Yurke85,La Porta89}.
Other proposals relied on feedforward modulation of signal modes and
off-line squeezed probe modes \cite{Buchler99,Andersen02,Filip05,Yoshikawa08},
in which it is not necessary to strongly pump a nonlinear medium in
line to the signal modes. The combination of off-line squeezed probe
modes, linear optics, homodyne detection, and feedforward loop has
inspired many other optical operations, such as squeezing \cite{Filip05,Buchler01,Yoshikawa07,Miyata14},
implementation of the one-way computation \cite{Ukai11,Yokoyama15},
realization of third-order nonlinear operation \cite{Marek11}, and
other varieties of QND interactions \cite{Marek10,Yokoyama14}.

In this paper we propose a nondestructive method to verify continuous-variable
bipartite entanglement, which uses QND and beam-splitter interactions
between the signal modes and two other probe modes. After the interactions,
the probe modes are measured by homodyne detections, so that the obtained
photocurrents serve both to calculate the entanglement condition,
so as to modulate the signal modes by electro-optic feedforward modulation,
as has been implemented in noiseless optical amplifiers \cite{Ralph97,Lam97,Huntington98}.
The proposed scheme has the benefit of not mixing up the output signal
modes to each other, which would change the global properties of entanglement
\cite{Simon,DGCZ}. Another relevant result is that the obtained entanglement
condition is valid to the output signals, ensuring the quantum correlation
properties resulting from the process. The cost of the procedure,
manageable by the scheme parameters, is the addition of excess uncorrelated
noise in both signal modes. 

This paper is organized as follows. In Section II, we describe the
entanglement verification procedure, in which is shown the quadrature
transformations of the signal and probe modes, in each stage of the
scheme. In Section III, we characterize the procedure by the well-known
QND measurement criteria. In Section IV, based on the results of the
previous section, we present how we compute a sufficient entanglement
condition to the two output signal modes, by detecting both probe
modes. The effects of the noise addition on the signal entanglement
are assessed in Section V, where the entanglement degradation is obtained
in the case of Gaussian systems. Finally, we discuss the results and
possibilities of this scheme in Section VI.

\section{Entanglement verification}

The goal of the process is to verify if a pair of signal modes are
entangled without destroying them. Since we consider the signal modes
as continuous-variable systems, we write their input quadrature operators
as $\hat{x}_{1}$ and $\hat{p}_{1}$ for mode 1 and $\hat{x}_{2}$
and $\hat{p}_{2}$ for mode 2. In order to carry out the entanglement
verification, two independent auxiliary beams must be introduced,
characterized with input quadrature operators $\hat{x}_{A}$ and $\hat{p}_{A}$
for mode A, and $\hat{x}_{B}$ and $\hat{p}_{B}$ for mode B. All
operators obey the usual commutation relations, $[\hat{x}_{i},\hat{p}_{j}]=i\delta_{ij}$,
for $\{i;j\}=\{1;2;A;B\}$ \cite{Braunstein05}. The auxiliary beams
are used as probe modes, interacting with the signal modes and after
measured by homodyne detections. Thus, the states of these probe modes
A and B must be previously known. Following the theoretical proposals
of previous articles \cite{Buchler99,Andersen02,Filip05}, the probe
modes must be prepared in strongly squeezed vacuum states. In what
follows, the squeezed quadratures of the probe modes are related to
the operators $\hat{p}_{A}$ and $\hat{x}_{B}$. First, each signal
mode is coupled to each probe mode by ideal QND interactions. Interactions
such as these have been performed by coupling beams in nonlinear optical
media (see \cite{Grangier98} and citations therein). However QND
interactions were also performed using only linear optics, an auxiliary
squeezed beam, homodyne detection, and feedforward modulation \cite{Filip05,Yoshikawa08}.
As illustrated in Figure 1, the beam pairs (1, A) and (2, B) are coupled
by QND interactions, such that for modes 1 and A, with gain $G_{1}$:
\begin{eqnarray}
\hat{x}_{1}^{\prime} & = & \hat{x}_{1},\label{QND-X1}\\
\hat{p}_{1}^{\prime} & = & \hat{p}_{1}-G_{1}\hat{p}_{A},\label{QND-P1}\\
\hat{x}_{A}^{\prime} & = & \hat{x}_{A}+G_{1}\hat{x}_{1},\label{QND-XA}\\
\hat{p}_{A}^{\prime} & = & \hat{p}_{A},\label{QND-PA}
\end{eqnarray}
and for modes 2 and B, with gain $G_{2}$: 
\begin{eqnarray}
\hat{x}_{2}^{\prime} & = & \hat{x}_{2}+G_{2}\hat{x}_{B},\label{QND-X2}\\
\hat{p}_{2}^{\prime} & = & \hat{p}_{2},\label{QND-P2}\\
\hat{x}_{B}^{\prime} & = & \hat{x}_{B},\label{QND-XB}\\
\hat{p}_{B}^{\prime} & = & \hat{p}_{B}-G_{2}\hat{p}_{2}.\label{QND-PB}
\end{eqnarray}

\begin{figure}
\begin{centering}
\includegraphics[width=8cm]{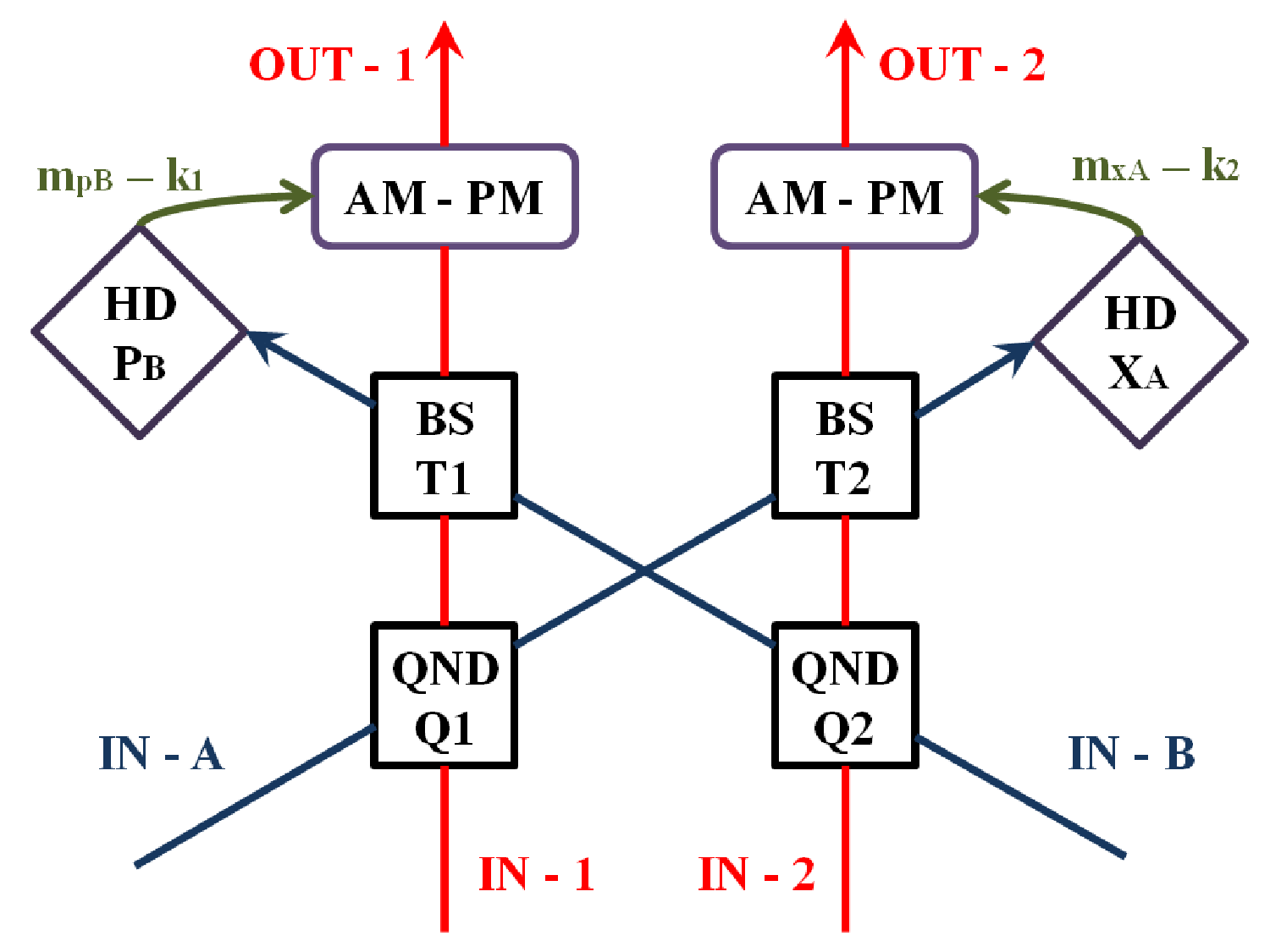} 
\par\end{centering}

\caption{Schematic setup for the entanglement verification of modes 1 and 2.
The probe modes are modes A and B. HD: homodyne detection for quadrature
operators $\hat{x}_{A}$ and $\hat{p}_{B}$; BS: beam-splitter coupling
with transmittances $T_{1}$ and $T_{2}$; QND: quantum nondemolition
coupling with gains $Q_{1}$ and $Q_{2}$; AM-PM: feedforward modulators.
$m_{x_{A}}$ and $m_{p_{B}}$ are the photocurrents obtained in detectors.
These signals are used in both the feedforward process and to compute
the entanglement condition.}
\end{figure}

After these first two QND interactions, the modes interact again,
crossing probe modes with signal modes. Both interactions operate
as beam splitters with transmittance $T_{1}$ for modes 1 and B and
transmittance $T_{2}$ for modes 2 and A. Thus from equations (\ref{QND-X1})-(\ref{QND-PB}),
we have to modes 1 and B: 
\begin{eqnarray}
\hspace{-2em}\hat{x}_{1}^{\prime\prime} & = & \sqrt{T_{1}}\hat{x}_{1}-\sqrt{1-T_{1}}\hat{x}_{B},\label{BS-X1}\\
\hspace{-2em}\hat{p}_{1}^{\prime\prime} & = & \sqrt{T_{1}}(\hat{p}_{1}-G_{1}\hat{p}_{A})-\sqrt{1-T_{1}}(\hat{p}_{B}-G_{2}\hat{p}_{2}),\label{BS-P1}\\
\hspace{-2em}\hat{x}_{B}^{\prime\prime} & = & \sqrt{T_{1}}\hat{x}_{B}+\sqrt{1-T_{1}}\hat{x}_{1},\label{BS-XB}\\
\hspace{-2em}\hat{p}_{B}^{\prime\prime} & = & \sqrt{T_{1}}(\hat{p}_{B}-G_{2}\hat{p}_{2})+\sqrt{1-T_{1}}(\hat{p}_{1}-G_{1}\hat{p}_{A}),\label{BS-PB}
\end{eqnarray}
and to modes 2 and A: 
\begin{eqnarray}
\hspace{-2em}\hat{x}_{2}^{\prime\prime} & = & \sqrt{T_{2}}(\hat{x}_{2}+G_{2}\hat{x}_{B})-\sqrt{1-T_{2}}(\hat{x}_{A}+G_{1}\hat{x}_{1}),\label{BS-X2}\\
\hspace{-2em}\hat{p}_{2}^{\prime\prime} & = & \sqrt{T_{2}}\hat{p}_{2}-\sqrt{1-T_{2}}\hat{p}_{A},\label{BS-P2}\\
\hspace{-2em}\hat{x}_{A}^{\prime\prime} & = & \sqrt{T_{2}}(\hat{x}_{A}+G_{1}\hat{x}_{1})+\sqrt{1-T_{2}}(\hat{x}_{2}+G_{2}\hat{x}_{B}),\label{BS-XA}\\
\hspace{-2em}\hat{p}_{A}^{\prime\prime} & = & \sqrt{T_{2}}\hat{p}_{A}+\sqrt{1-T_{2}}\hat{p}_{2}.\label{BS-PA}
\end{eqnarray}

These first two steps are necessary for the probe modes to obtain
sufficient information from the signal modes. Then the probe modes
are measured by homodyne detection processes, providing classical
signals (photocurrents) sufficient to compute and verify if the signal
modes are entangled. However, as we see in equations (\ref{BS-X1}),
(\ref{BS-P1}), (\ref{BS-X2}) and (\ref{BS-P2}), the signal modes
are affected by interactions. These perturbations can be corrected
\textit{a posteriori} with feedforward modulations, using phase and
amplitude electro-optic modulators. Setting up the local oscillators
of the homodyne detections, we select the quadratures $\hat{x}_{A}^{\prime\prime}$
and $\hat{p}_{B}^{\prime\prime}$ to measure. With this choice, we
obtain access to all signal quadrature operators, found in linear
combinations in equations (\ref{BS-PB}) and (\ref{BS-XA}). In Section
IV, we show such measures are sufficient for the entanglement verification. 

As the currently available detectors can achieve efficiencies above
99\% \cite{Yoshikawa08}, we will discard the noise from the detection
process, so that we will focus only on the inherent aspects of the
procedure. However, the noise from detector imperfections can be calculated,
which would add vacuum fluctuation terms in our derivations. 

The photocurrents generated in the detectors, $m_{p_{B}}$ and $m_{x_{A}}$,
are amplified with gains $k_{1}$ and $k_{2}$ for electro-optic modulators,
so that the following signal quadratures are transformed as 
\begin{equation}
\hat{p}_{1}^{\prime\prime\prime}=\hat{p}_{1}^{\prime\prime}+k_{1}m_{p_{B}}\label{FF-P1}
\end{equation}
and 
\begin{equation}
\hat{x}_{2}^{\prime\prime\prime}=\hat{x}_{2}^{\prime\prime}+k_{2}m_{x_{A}}.\label{FF-X2}
\end{equation}
The respective conjugate quadratures remain unchanged. The gains $k_{1}$
and $k_{2}$ must be tuned in a way that the crossed terms between
modes 1 and 2 are canceled.

In conclusion, we can implement squeezing operations with gains $T_{1}$
and $T_{2}$ onto signal modes 1 and 2 (not shown in Figure 1) , so
that we obtain the output modes 
\begin{eqnarray}
\hat{x}_{1}^{out} & = & \hat{x}_{1}-g_{1}\hat{x}_{B},\label{OUT-X1}\\
\hat{p}_{1}^{out} & = & \hat{p}_{1}-G_{1}\hat{p}_{A},\label{OUT-P1}\\
\hat{x}_{2}^{out} & = & \hat{x}_{2}+G_{2}\hat{x}_{B},\label{OUT-X2}\\
\hat{p}_{2}^{out} & = & \hat{p}_{2}-g_{2}\hat{p}_{A},\label{OUT-P2}
\end{eqnarray}
where $g_{1}=\sqrt{(1-T_{1})/T_{1}}$ and $g_{2}=\sqrt{(1-T_{2})/T_{2}}$.
The last squeezing operations onto modes 1 and 2 are not critical,
because the entanglement is invariant under local linear unitary Bogoliubov
operations \cite{Simon,DGCZ}. In equations (\ref{OUT-X1})-(\ref{OUT-P2}),
we maintain the terms with strongly squeezed quadratures, $\hat{p}_{A}$
and $\hat{x}_{B}$, although their variances tend to vanishing. At
this point, it is interesting to present every possible resulting
noise inherent to the scheme, reminding one that losses due to the
efficiencies of the detectors are not being considered. In fact, the
variances of $\hat{p}_{A}$ and $\hat{x}_{B}$ are smaller the larger
the squeezing in the probe modes. Nevertheless we need to know the
scales of $g_{i}$ and $G_{i}$, $i=\{1,2\}$, before ruling out negligible
terms. In the next section, the conditions to a genuine QND process
will be studied, in which it is shown that the parameters $g_{i}$
and $G_{i}$ can be tuned to optimize the scheme, so that some terms
are eventually negligible.

\section{QND Characterization}

As we are studying a procedure that must conserve some property of
the signal modes, we must check it regarding the features of a QND
process. In early articles on QND measurement in optical systems,
quantities were settled to characterize a device if it works as a
noiseless amplifier and as a quantum state preparation (QSP) \cite{Grangier98,Imoto89,Holland90,Blockley90,Grangier92,Roch92}.
To assess these features, we must consider quantities connecting statistical
properties of the input and output modes, to both signal and probe
pairs. The noise inserted in the system can be quantified by signal-to-noise
ratios of the input signal, $R_{s}^{in}$, output signal, $R_{s}^{out}$,
and output probe, $R_{p}^{out}$. The transfer coefficients from the
input signal to the output signal are given by 
\begin{equation}
\mathcal{T}_{s}=\frac{R_{s}^{out}}{R_{s}^{in}}=\frac{V_{s}^{in}}{V_{s}^{in}+N_{s}},\label{transfer-signal}
\end{equation}
 and from the input signal to the output probe by 
\begin{equation}
\mathcal{T}_{p}=\frac{R_{p}^{out}}{R_{s}^{in}}=\frac{V_{s}^{in}}{V_{s}^{in}+N_{p}},\label{transfer-probe}
\end{equation}
 where $V_{s}^{in}$ is the input signal quadrature variance, and
$N_{s}$ and $N_{p}$ are the equivalent input noises related to signal
and probe inputs, respectively. Another quantity of interest is the
conditional variance of the output signal related to the measured
output probe, given by 
\begin{equation}
W_{QSP}=V_{s}^{out}-\frac{|C_{s,p}^{out}|^{2}}{V_{p}^{out}},\label{conditional-variance}
\end{equation}
 where $V_{s}^{out}$ and $V_{p}^{out}$ are the signal and probe
output quadrature variances, respectively, and $C_{s,p}^{out}$ is
the symmetrized covariance between former quadratures. According to
early articles \cite{Imoto89,Holland90,Blockley90,Grangier92,Roch92},
a fully QND process must simultaneously meet the following conditions:
\begin{equation}
\mathcal{T}_{s}+\mathcal{T}_{p}>1,\label{transfer-condition}
\end{equation}
indicating the noiseless amplifier property (quantum optical tapping),
and 
\begin{equation}
W_{QSP}<1,\label{variance-condition}
\end{equation}
 indicating the quantum state preparation property. 

In the case of the entanglement verification, there are two output
signals, each one with two quadratures, and two output probe quadratures.
So we must calculate transfer coefficients (\ref{transfer-signal})
and (\ref{transfer-probe}) and conditional variances (\ref{conditional-variance})
to a bipartite signal and a bipartite probe. That entails more combinations
among the quadrature operators, implying more transfer coefficients
and conditional variances to be considered. At first, we can seek
for all combinations of quadrature operators between signal and probe
systems. On the other hand, transfer coefficients crossing quadratures
do not exist (for example, there is no $N_{p_{B}}$ related to $\hat{x}_{1}$).
Moreover, a direct verification unveils that conditions (\ref{transfer-condition})
and (\ref{variance-condition}) cannot be satisfied in all existing
combinations of quadrature operators. However, to meet a quantum regime,
it is sufficient to comply only with conditions (\ref{transfer-condition})
and (\ref{variance-condition}) related with the signal quadratures
preserved in the QND procedure. Therefore, all these considerations
restrict the relevant transfer coefficients and conditional variances.
For example, we can choose the signal quadrature operators $\hat{x}_{1}$
and $\hat{p}_{2}$ to be preserved. Thus the equivalent noises, introduced
in systems, are 
\begin{eqnarray}
\hspace{-2em}N_{x_{1}} & = & g_{1}^{2}\langle(\Delta\hat{x}_{B})^{2}\rangle,\label{equiv-noise-X1}\\
\hspace{-2em}N_{p_{2}} & = & g_{2}^{2}\langle(\Delta\hat{p}_{A})^{2}\rangle,\label{equiv-noise-P2}\\
\hspace{-2em}N_{p_{B}}^{(p_{2})} & = & \frac{1}{G_{2}^{2}}\langle(\Delta\hat{p}_{B})^{2}\rangle+\left(\frac{g_{1}}{G_{2}}\right)^{2}\langle(\Delta\hat{p}_{1})^{2}\rangle\nonumber \\
\hspace{-2em} & - & \frac{g_{1}}{G_{2}}\langle\tfrac{1}{2}\{\Delta\hat{p}_{1},\Delta\hat{p}_{2}\}\rangle+\left(\frac{g_{1}G_{1}}{G_{2}}\right)^{2}\langle(\Delta\hat{p}_{A})^{2}\rangle,\label{equiv-noise-PB}\\
\hspace{-2em}N_{x_{A}}^{(x_{1})} & = & \frac{1}{G_{1}^{2}}\langle(\Delta\hat{x}_{A})^{2}\rangle+\left(\frac{g_{2}}{G_{1}}\right)^{2}\langle(\Delta\hat{x}_{2})^{2}\rangle\nonumber \\
\hspace{-2em} & + & \frac{g_{2}}{G_{1}}\langle\tfrac{1}{2}\{\Delta\hat{x}_{1},\Delta\hat{x}_{2}\}\rangle+\left(\frac{g_{2}G_{2}}{G_{1}}\right)^{2}\langle(\Delta\hat{x}_{B})^{2}\rangle,\label{equiv-noise-XA}
\end{eqnarray}
where $\Delta\hat{\mathcal{O}}_{i}=\hat{\mathcal{O}}_{i}-\langle\hat{\mathcal{O}}_{i}\rangle$
and $\langle\hat{\mathcal{O}}_{i}\rangle={\rm Tr}(\hat{\mathcal{O}}_{i}\hat{\rho})$,
such that $\hat{\mathcal{O}}_{i}$ is some operator distinguished
by index $i$, and $\hat{\rho}$ is the density matrix of the whole
system. With expressions (\ref{equiv-noise-X1})-(\ref{equiv-noise-XA}),
we can find the respective conditions (\ref{transfer-condition}).
After simple algebra, such conditions are rewritten as 
\begin{equation}
N_{x_{1}}N_{x_{A}}^{(x_{1})}<\langle(\Delta\hat{x}_{1})^{2}\rangle^{2}\label{equiv-condition-X}
\end{equation}
 and 
\begin{equation}
N_{p_{2}}N_{p_{B}}^{(p_{2})}<\langle(\Delta\hat{p}_{2})^{2}\rangle^{2}.\label{equiv-condition-P}
\end{equation}
So it is clear that sufficiently small values of $g_{1}$ and $g_{2}$
and sufficiently large values of $G_{1}$ and $G_{2}$ will achieve
a quantum optical tapping regime. The ideal situation is obtained
when $g_{1};g_{2}\rightarrow0$ and $G_{1};G_{2}\rightarrow\infty$,
in which perfectly noiseless amplifiers are achieved. 

Conditional variance (\ref{conditional-variance}) applied to output
modes unfolds in two other quantities: 
\begin{equation}
W_{QSP}(\hat{x}_{1},\hat{x}_{A})=\langle(\Delta\hat{x}_{1}^{out})^{2}\rangle-\frac{|\langle\tfrac{1}{2}\{\Delta\hat{x}_{1}^{out},\Delta\hat{x}_{A}^{out}\}\rangle|^{2}}{\langle(\Delta\hat{x}_{A}^{out})^{2}\rangle}\label{conditional-variance-X}
\end{equation}
and 
\begin{equation}
W_{QSP}(\hat{p}_{2},\hat{p}_{B})=\langle(\Delta\hat{p}_{2}^{out})^{2}\rangle-\frac{|\langle\tfrac{1}{2}\{\Delta\hat{p}_{2}^{out},\Delta\hat{p}_{B}^{out}\}\rangle|^{2}}{\langle(\Delta\hat{p}_{B}^{out})^{2}\rangle}.\label{conditional-variance-P}
\end{equation}
It is possible to find conditions to the quantum state preparation
regime with both expressions (\ref{conditional-variance-X}) and (\ref{conditional-variance-P})
simultaneously. A limit case can be obtained, considering strongly
squeezed input probe modes, such that $\langle(\Delta\hat{x}_{B})^{2}\rangle;\langle(\Delta\hat{p}_{A})^{2}\rangle\rightarrow0$,
and based on previous considerations, taking $g_{1};g_{2}\rightarrow0$,
we notice that the quantum state preparation is attainable if, from
expression (\ref{conditional-variance-X}), 
\begin{equation}
G_{1}^{2}>\left(1-\frac{1}{\langle(\Delta\hat{x}_{1})^{2}\rangle}\right)\langle(\Delta\hat{x}_{A})^{2}\rangle.\label{variance-condition-X1-A}
\end{equation}
As $(1-1/\langle(\Delta\hat{x}_{1})^{2}\rangle)<1$, to any physical
value of $\langle(\Delta\hat{x}_{1})^{2}\rangle$, a stricter inequality
is more interesting, 
\begin{equation}
G_{1}^{2}\geq\langle(\Delta\hat{x}_{A})^{2}\rangle.\label{variance-condition-XA}
\end{equation}
Similarly, from expression (\ref{conditional-variance-P}), we can
obtain another inequality, 
\begin{equation}
G_{2}^{2}\geq\langle(\Delta\hat{p}_{B})^{2}\rangle.\label{variance-condition-PB}
\end{equation}
Both inequalities (\ref{variance-condition-XA}) and (\ref{variance-condition-PB})
can be fulfilled simultaneously, therefore quantum state preparation
regimes are feasible for reasonable values of the parameters of the
optical device, independently of the input beam properties.

Considering other combinations of signal quadrature operators to be
preserved in the QND procedure, we can seek other conditions to $g_{1}$,
$g_{2}$, $G_{1}$, and $G_{2}$, analogously to previous analysis.
Such cases are very similar and do not add new information. For the
example studied, we can safely approximate the output modes to 
\begin{eqnarray}
\hat{x}_{1}^{out} & = & \hat{x}_{1},\label{OUT-X1-B}\\
\hat{p}_{1}^{out} & = & \hat{p}_{1}-G_{1}\hat{p}_{A},\label{OUT-P1-B}\\
\hat{x}_{2}^{out} & = & \hat{x}_{2}+G_{2}\hat{x}_{B},\label{OUT-X2-B}\\
\hat{p}_{2}^{out} & = & \hat{p}_{2}.\label{OUT-P2-B}
\end{eqnarray}
 We can notice in the output signals that each mode has a preserved
quadrature, featuring a QND process. On the other hand, each mode
has a conjugate quadrature added by terms from probe modes. As the
input probe modes are independent, these terms produce phase-sensitive
uncorrelated noise, inevitably disturbing the signal modes.

\section{Calculating the entanglement condition }

Besides feedforward modulation, the photocurrents are also used to
calculate the entanglement condition of the signal modes. Duan \textit{et
al.} \cite{DGCZ} have found a sufficient entanglement condition in
continuous-variable systems, based on EPR-like operators. Later, other
works have extended this condition for more general operators \cite{Simon01,Giovannetti03}.
Following these authors, consider operator combinations such as 
\begin{eqnarray}
\hat{u} & = & a_{1}\hat{x}_{1}+a_{2}\hat{x}_{2},\\
\hat{v} & = & b_{1}\hat{p}_{1}+b_{2}\hat{p}_{2},
\end{eqnarray}
where $[\hat{x}_{i},\hat{p}_{j}]=i\delta_{ij}$, ($i;j=1;2$), and
$a_{1}$, $a_{2}$, $b_{1}$, and $b_{2}$ are arbitrary constants.
It is possible to show that a sufficient condition of continuous-variable
bipartite entanglement is 
\begin{equation}
\langle(\Delta\hat{u})^{2}\rangle+\langle(\Delta\hat{v})^{2}\rangle<|a_{1}b_{1}|+|a_{2}b_{2}|,\label{entangle-condition}
\end{equation}
 valid to Gaussian or non-Gaussian systems.

On the other hand, the variances of the photocurrents generated in
the homodyne detection can be written as 
\begin{eqnarray}
\hspace{-2em}\langle(\Delta m_{X_{A}})^{2}\rangle & = & K_{A}T_{2}\left[\langle(\Delta\hat{u}^{out})^{2}\rangle+\langle(\Delta\hat{x}_{A})^{2}\rangle\right],\label{photocurrent-A}\\
\hspace{-2em}\langle(\Delta m_{P_{B}})^{2}\rangle & = & K_{B}T_{1}\left[\langle(\Delta\hat{v}^{out})^{2}\rangle+\langle(\Delta\hat{p}_{B})^{2}\rangle\right],\label{photocurrent-B}
\end{eqnarray}
where $\hat{u}^{out}=G_{1}\hat{x}_{1}^{out}+g_{2}\hat{x}_{2}^{out}$
and $\hat{v}^{out}=g_{1}\hat{p}_{1}^{out}-G_{2}\hat{p}_{2}^{out}$
are defined from equations (\ref{OUT-X1-B}) to (\ref{OUT-P2-B}).
$K_{A}$ and $K_{B}$ are factors that depend on the overall detector
efficiencies and the conversion circuitry of the photocurrents, such
that these factors can be related by $K_{A}=(g_{2}/k_{2})^{2}$ and
$K_{B}=(g_{1}/k_{1})^{2}$. Except for $\langle(\Delta\hat{u}^{out})^{2}\rangle$
and $\langle(\Delta\hat{v}^{out})^{2}\rangle$, all other terms of
equations (\ref{photocurrent-A}) and (\ref{photocurrent-B}) are
measured or previously known. So both $\langle(\Delta\hat{u}^{out})^{2}\rangle$
and $\langle(\Delta\hat{v}^{out})^{2}\rangle$ can be calculated and
compared with equation (\ref{entangle-condition}), identifying the
arbitrary constants with the parameters of the apparatus, i. e., $a_{1}=G_{1}$,
$a_{2}=g_{2}$, $b_{1}=g_{1}$, and $b_{2}=-G_{2}$. Therefore the
entanglement condition (\ref{entangle-condition}) can be calculated
from the detected probe modes and from parameters of the scheme. We
can also notice that, to sufficiently squeezed probe modes, the photocurrent
variances are directly proportional to the EPR-like operator variances,
i. e., $\langle(\Delta m_{X_{A}})^{2}\rangle\simeq K_{A}T_{2}\langle(\Delta\hat{u}^{out})^{2}\rangle$
and $\langle(\Delta m_{P_{B}})^{2}\rangle\simeq K_{B}T_{1}\langle(\Delta\hat{v}^{out})^{2}\rangle$,
so that the photocurrent measurements provide a direct way to verify
the entanglement. 

An important result of this method is that the condition of entanglement
(\ref{entangle-condition}), calculated with expressions (\ref{photocurrent-A})
and (\ref{photocurrent-B}), is exactly valid for the output signal
modes, namely, we can certify that the signals resulting from the
scheme are entangled, regardless of limitations or scheme losses.
From a practical point of view, we are receiving two signal modes
to verify the quantum properties of their correlations, so that we
can nondestructively maintain them for use in other processes, and
still be able to repeat it. Inevitably the scheme has a cost, which
is the addition of phase-sensible noise, degrading the signal modes.
The effects of this degradation on the entanglement are discussed
in the next section. In an idealized situation, the probe modes would
have a squeezing parameter tending to infinity, so the signal modes
could have their entanglement checked without any degradation, perfectly
preserving each quadrature of the signals as well.

\section{Entanglement degradation}

To assess the effects that the presented scheme cause on the signal
mode entanglement, we will restrict this analysis to the case of Gaussian
beams \cite{Weedbrook12}. The continuous-variable systems restricted
to Gaussian states are fully described by the first statistical moments
of the dynamical operators, i. e., $\mathcal{\bar{O}}_{i}\equiv\langle\hat{\mathcal{O}}_{i}\rangle$,
and by the second statistical moments, that can be arranged in a covariance
matrix, $M$, whose entries are $M_{ij}\equiv\tfrac{1}{2}\langle\{\Delta\hat{\mathcal{O}}_{i},\Delta\hat{\mathcal{O}}_{j}\}\rangle$.
With a suitable choice of quadrature basis, we can write the input
covariance matrix as 
\begin{equation}
M^{in}=\left(\begin{array}{cccc}
n_{1} & 0 & c & 0\\
0 & n_{1} & 0 & k\\
c & 0 & n_{2} & 0\\
0 & k & 0 & n_{2}
\end{array}\right).\label{M-in}
\end{equation}
After all procedures, the covariance matrix of the signal modes becomes
\begin{equation}
M^{out}=\left(\begin{array}{cccc}
n_{1} & 0 & c & 0\\
0 & n_{1}+d_{1} & 0 & k\\
c & 0 & n_{2}+d_{2} & 0\\
0 & k & 0 & n_{2}
\end{array}\right),\label{M-out}
\end{equation}
in which we can observe the presence of uncorrelated excess noises
$d_{1}=G_{1}^{\phantom{}2}\langle(\Delta\hat{p}_{A})^{2})\rangle$
and $d_{2}=G_{2}^{\phantom{}2}\langle(\Delta\hat{x}_{B})^{2})\rangle$,
generated from probe operators in equations (\ref{OUT-P1-B}) and
(\ref{OUT-X2-B}). These noises spoil the input entanglement. This
can be seen by calculating the logarithmic negativity \cite{Vidal02,Adesso04,Plenio05}:
\begin{equation}
E_{N}(\rho)=\max[0,-\mathrm{Ln}\tilde{\nu}_{-}],\label{Log-Neg}
\end{equation}
where $\tilde{\nu}_{-}$ is the smallest symplectic eigenvalue of
the partially transposed bipartite state. This quantity can be calculated
from symplectic invariants of the partially transposed system: 
\begin{equation}
\tilde{\nu}_{-}=\sqrt{\frac{\tilde{\Delta}-\sqrt{\tilde{\Delta}^{2}-4\det M}}{2}},\label{simpletic-eigenvalue}
\end{equation}
where $\tilde{\Delta}=n_{1}(n_{1}+d_{1})+n_{2}(n_{2}+d_{2})-2ck$
and $\det M=[n_{1}(n_{2}+d_{2})-c^{2}][n_{2}(n_{1}+d_{1})-k^{2}]$,
if we use matrix (\ref{M-out}). Hence we find the expression very
complex. To simplify the problem, we consider input signal modes as
two-mode squeezed vacuum, so $n_{1}=n_{2}=\cosh(2r)$ and $c=-k=\sinh(2r)$.
With these substitutions, we plot Figure 2. 

\begin{figure}[H]
\begin{centering}
\includegraphics[width=8cm]{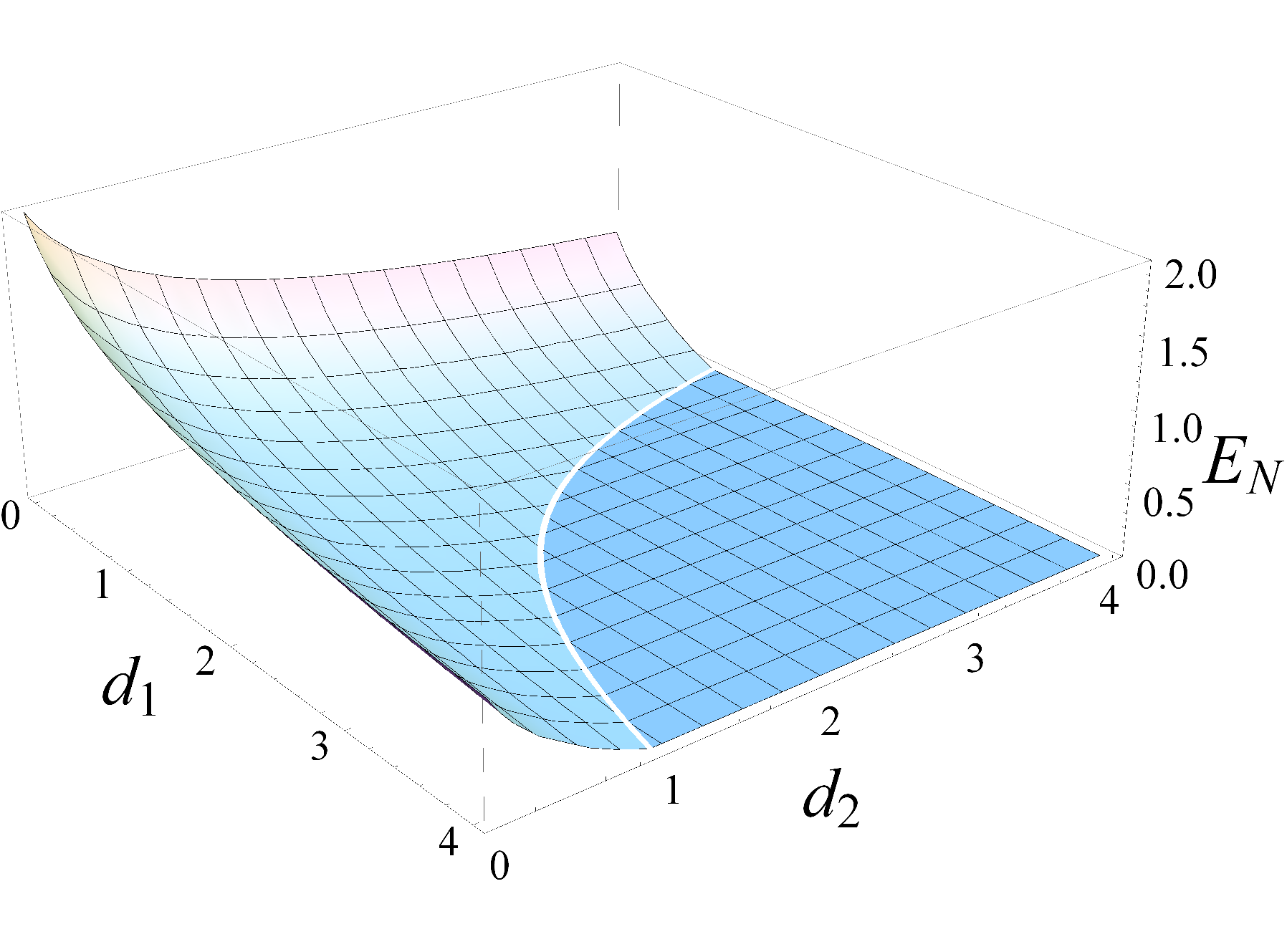} 
\par\end{centering}

\caption{Plot of logarithmic negativity to a two-mode Gaussian state characterized
by matrix (\ref{M-out}) as the function of excess noises $d_{1}$
and $d_{2}$. The input modes are two-mode squeezed pure states, with
squeezing parameter $r=1$. }
\end{figure}

One can see in Figure 2 that, for any nonzero values of $d_{1}$ and
$d_{2}$, the logarithmic negativity is monotonically decreasing.
Such effect is the verification process cost. The optimization is
reached minimizing the variances of $\hat{p}_{A}$ and $\hat{x}_{B}$,
i. e., increasing the squeezing of the probe modes. In addition, the
gains $G_{1}$ and $G_{2}$ must be limited, obeying conditions (\ref{equiv-condition-X}),
(\ref{equiv-condition-P}), (\ref{variance-condition-XA}), and (\ref{variance-condition-PB}).

\section{Discussion}

In this paper, a nondestructive scheme for bipartite entanglement
verification in the framework of continuous-variable systems was shown.
To such task, a suitable choice of QND and beam-splitter interactions
between the pair of signal modes and a pair of probe modes is necessary,
followed by measurements of the probe modes and feedforward modulations
of the signal modes. All of these processes are feasible with current
technologies, therefore it can be performed in experimental demonstrations
or implemented as a built-in step in a larger communication protocol,
in which it is necessary to certify that two signals are entangled,
while they are used in another further step. Some studies have already
been done with similar purposes in the context of discrete variable
systems \cite{Barrett05,Tang12,Liu16}. So this paper fills a gap
for continuous-variable systems.

This method is based on a sufficient entanglement condition, therefore
some entangled states cannot be detected. However, entangled states
that do not satisfy inequality (\ref{entangle-condition}) are fragile
when subjected to Gaussian attenuation, as already shown in the articles
by Barbosa \textit{et al.} \cite{Barbosa10,Barbosa11}. So these fragile
states would not be interesting for long-distance implementations.
Moreover, maximally entangled states or near them are required for
many quantum information protocols. Such states are addressed by the
scheme.

According to the criteria of QND measurement characterization \cite{Imoto89,Holland90,Blockley90,Grangier92,Roch92},
we assess the quantum properties of the entanglement verification
scheme. One notice that there are many possibilities to calculate
the transfer coefficients and the conditional variances, using different
quadratures of the bipartite modes. In this paper, we calculate all
relevant quantities to certify the QND properties, although we do
not present a systematic method for finding them. Therefore a QND
multipartite device characterization would be a relevant theoretical
development for further research. 

The entanglement of the signal modes can be checked by the presented
method, but it cannot quantify the entanglement. Such limitation exists
because the photocurrents, measured by homodyne detectors, provide
information only about the variances of the EPR-like operators, $a_{1}\hat{x}_{1}+a_{2}\hat{x}_{2}$
and $b_{1}\hat{p}_{1}+b_{2}\hat{p}_{2}$. That is insufficient to
have a measure of entanglement, e. g., logarithmic negativity, although
it is sufficient to detect entanglement. However, new strategies may
lead to quantifying the entanglement. We can expect that more complex
signal-probe interaction configurations reach these goals. Unlike
squeezed vacuum modes, using non-Gaussian modes as probe modes could
also lead to more promising results, as already noted in articles
concerned with the trade-off between information and disturbance caused
by measurements in continuous-variable systems \cite{Andersen06,Mista06}.
All these considerations show many future research possibilities. 

We can notice that the entanglement verification is deeply connected
with the eavesdropping in quantum cryptography \cite{Gisin02}. While
an eavesdropper wants to make a necessarily imperfect copy of the
signal sent between two communication stations, in the proposed entanglement
verification, it looks to observe the correlation between the two
signal modes, without necessarily copying and measuring the physical
states. The similarity between the two processes is that both are
extracting information from the signal, and both are adding unavoidable
perturbation, in the case of this paper, adding uncorrelated phase-sensible
noise. In both processes, the minimization of the perturbation is
the condition to its optimal accomplishment. Further studies may be
devoted to the details and explanations of the relationship between
the QND verification of multipartite correlations and cloning and
eavesdropping.
\begin{acknowledgments}
The author acknowledges support from FAPEMIG (Fundação de Amparo à
Pesquisa do Estado de Minas Gerais), Grant No. CEX-APQ-01899-13. \end{acknowledgments}

\end{document}